\documentclass[useAMS,usenatbib]{mn2e}
\usepackage{graphicx}

\usepackage{graphics,color}
\definecolor{dark}{gray}{0.5}
\definecolor{red}{rgb}{1,0,0}
\definecolor{green}{rgb}{0,1,0}
\definecolor{blue}{rgb}{0,0,1}

\title{Survivability of a star cluster in a dispersing molecular cloud}
\author[H.C. Chen and C.M. Ko]{Hui-Chen Chen$^{1}$
\thanks{E-mail: huichen@astro.ncu.edu.tw (HCC);
cmko@astro.ncu.edu.tw (CMK)} and Chung-Ming Ko$^{1,2}\footnotemark[1]$ \\
$^1$  Institute of Astronomy, National Central University, Taiwan \\
$^2$  Department of Physics and Center for Complex Systems, National Central University,
Taiwan}

\begin{document}

\date{Accepted . Received ; in original form 2008 April ?}

\pagerange{\pageref{firstpage}--\pageref{lastpage}} \pubyear{2002}

\maketitle

\label{firstpage}

\begin{abstract}
Star clusters are formed in molecular clouds which are believed
to be the birth places of most stars. From recent observational data,
\cite{Lada2003} estimated that only 4\% to 7\% of the
clusters embedded inside molecular clouds have survived.
An important mechanism for the disruption of embedded (bound)-clusters is the dispersion
of the parent cloud by UV radiation, stellar winds and/or supernova explosions.
In this work we study the effect of this mechanism by N-body simulations.
We find that most embedded-clusters survive for more than 30 Myr even when
different initial conditions of the cluster may introduce some minor variations,
but the general result is rather robust.

\end{abstract}


\section{Introduction}
Stellar clusters are among the most interesting objects in astronomy.
We are interested in how they form, how they evolve and how they die.
Observations gave the evidence that most stars are not born independently
but in stellar clusters or stellar associations
which are formed in molecular clouds (e.g., \cite{Lada2003}).
Due to the limits of
observational techniques, we do not know in detail
the relation between the clusters and the parent molecular clouds.
It is believed that stars are born in clusters and become field
stars after the clusters disassociate.

Recently, near infrared observational data (2MASS, Two Micron All Sky Survey
project at IPAC/Caltech) have shown that the number of embedded clusters is much
higher than the number of optical clusters for which the parent clouds
have already dissipated and that the survival probability for
embedded clusters in Milky Way is about 4\% to 7\% (\cite{Lada2003}).
Similar evidences for infant mortality are found in Antennae galaxies (\cite{Fall2005}),
Small Magellanic Cloud (\cite{Chandar2006}) and NGC1313 (\cite{Pellerin2007}).

These imply that clusters are likely to
be disrupted before the clouds are dissipated completely.
Since the time the clusters were born, they are under constant threats
from their surrounding environment.

Galactic tidal forces, close encounters with giant molecular
clouds (see, e.g., \cite{Gieles2006}), shock heating and mass loss
by massive member stars (see e.g., \cite{Boily2003a,Boily2003b})
are possible disruption mechanisms. Nonetheless, most of these
mechanisms have a destruction timescale longer than the upper
limit of the lifetime of molecular clouds which is about few to
few tens Myr (see e.g., \cite{Blitz1980}, \cite{Elmegreen2000},
\cite{Hartmann2001}, \cite{Bonnell2006}). The results of CO
observations in the Galaxy suggest that the lifetime of molecular
clouds is of the order of 10 Myr (see, e.g.,
\cite{Leisawitz1989}). The estimation of the timescale for
photoevaporation and statistics on the expected numbers of stars
per cloud show that giant molecular clouds of mass 10$^6$
M$_\odot$ are expected to survive for about 30 Myr (see, e.g.,
\cite{Williams1997}).

Generally speaking, mechanisms with destruction timescales less than the
lifetime of the clouds should be responsible for the low survival probability mentioned
above. In this work, we focus on the role played by the dispersion of the parent cloud on the
early evolution of the embedded clusters.

In the beginning, the clusters are bound to their parent molecular cloud.
As the cloud dissipates, the binding energy from the cloud decreases
and the stellar systems become out of equilibrium. Once out of equilibrium,
they may expand or dissociate completely.

The survivability of clusters under gas dispersion has been examined in the past
and extensive N-body simulations were performed in the past few years
(see e.g \cite{Lada1984}, \cite{Goodwin1997}, \cite{Boily2003b}, \cite{Baumgardt2007},
\cite{Bastian2006}, \cite{Goodwin2006}).
In a large set of simulations, \cite{Baumgardt2007} studied the dispersal of the residual gas by
decreasing the mass with different star formation efficiency (SFE),
and in different tidal fields.
They concluded that the clusters had to form with SFE $\geq$ 30\%
in order to survive gas expulsion, and the external tidal fields have significant influences only
if the ratio of half mass radius to tidal radius is larger than 0.05.
\cite{Goodwin2006} and \cite{Bastian2006} addressed similar problem and found that the embedded
clusters would be destroyed within a few tens of Myr if the ``effective star formation efficiency''
$\leq$ 30\% (eSFE$=1/2Q$, and Q=0.5 for virial equilibrium).


The paper is organized as follows.
In Section~2, we describe the model and parameters for our simulations.
In Section~3, we present and discuss the results and statistics of the simulations.
A summary and some remarks are provided in Section~4.

\section{Model and Simulation}

We intend to learn the behaviour of a star cluster in a dispersing molecular cloud.
For simplicity, we do not consider the feedback of the cluster onto the cloud, and the
cloud is simply represented by its gravitational force.
In other words, we study the behaviour of a star cluster in a time varying
gravitational field.
We adopted the N-body simulation code NBODY2 developed by \cite{Aarseth2001} for our
calculations.
As a first attempt, we assume a spherically symmetric external potential (representing the
molecular cloud), and initially the centre of mass of the cluster coincides with the
centre of the potential.
The initial spatial and velocity distributions of the cluster depend on the
initial profile of the external potential
(i.e., the initial mass and compactness of the cloud).
Subsequent evolution of the cluster depends, of course, on its initial distribution and
the rate of dispersion of the external potential.

\subsection{Model for the cloud}
The cloud is represented by a Plummer potential,

\begin{equation}
\Phi_{P}=\frac{-GM_b}{\sqrt{r^2+a^2}}\,,
\end{equation}
where $M_b$ is the total mass of the cloud, $a$ is the length
scale of the potential and $G$ is the gravitational constant.
To model the dispersion of the cloud we consider the potential to evolve
in time according to
\begin{equation}
a=a_0 e^{\alpha t}\,,
\end{equation}
where $a_0$ and $\alpha$ are the initial length scale and the dispersion rate
of the cloud, respectively.
The total mass of the cloud remains constant as the length scale
increases with time.
Stellar masses would change by stellar evolution.
However, since we run our simulations for 30 Myr only (which is about the maximum
life time of molecular clouds) and massive stars are rare,
hence we do not consider stellar evolution in this work.

There are three parameters for the cloud: (i) dispersion rate $\alpha$,
(ii) total mass $M_b$, and (iii) initial length scale $a_0$.

\subsubsection{Dispersion rate $\alpha$}
We take $\alpha$ as 0.1, 0.2, 0.3, 0.4, 0.5 in the unit used by the code.
These correspond to a e-fold time $t_e$ as 3.3, 1.5, 1.1, 0.75, 0.625 Myr in real unit
(see Table~\ref{para_time}).

\begin{table}
\caption{Dispersion rates of the cloud}
\begin{center}
\begin{tabular}{l c c c c c}
index  & A & B & C & D & E \\ \hline
 $\alpha$  & 0.1 & 0.2 & 0.3 & 0.4 & 0.5 \\
 $t_e$  [Myr] & 3.3 & 1.5 & 1.1 & 0.75 & 0.625 \\
\end{tabular}
\end{center}
\label{para_time}
\end{table}

\begin{figure}
\includegraphics[width=85mm, angle=0]{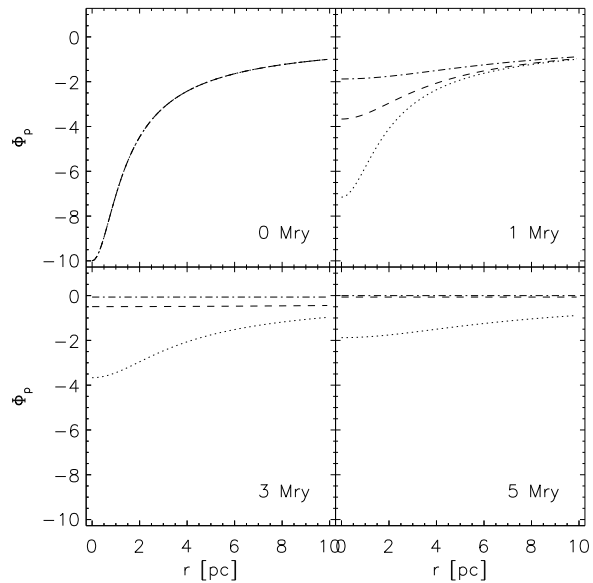}
\caption{
         Evolution of the cloud potential in dispersion rate of
         $\alpha$=0.1(dotted line),
         0.3 (dashed line) and 0.5 (dash-dotted line), whose e-fold times are
         $t_e$=3.3, 1.1 and 0.625 Myr, respectively.
         At 0 Myr, the potentials are the same.
         After 1 Myr, every one becomes shallower in difference degree.
         At 3 Myr, the two cases with $t_e$=1.1 and 0.625 Myr approach zero.
         At 5 Myr, even the case with $t_e$=3.3 Myr becomes rather flat.
         }
\label{plummerp}
\end{figure}

Fig.~\ref{plummerp} shows how the cloud potential evolves.
The potentials of $\alpha$ = 0.3 and 0.5 are almost zero after 5 Myr.
Even for $\alpha$ = 0.1, the potential is rather flat after 5 Myr.
We would expect the cloud exerts no effect on the cluster after
a relatively short time in these dispersion rates.
The seed of destruction is planted (if at all) only in the early dispersion stage
of the cloud.

\subsubsection{Total mass $M_b$}
The more massive the cloud is the more influence on the cluster is.
We consider the total mass of the cloud $M_b$ from 0.5 to 10 cluster mass $M_c$
as listed in Table~\ref{para_mass}.
We note that $M_c=2500M_\odot$ in all our simulations.
Generally, the star formation efficiency is defined as
\begin{equation}
\eta=\frac{M_{c}}{M_{c}+M_b}\,.
\end{equation}
The mass range we choose corresponds to $\eta=$ 50 to 9\%.

\begin{figure}
\includegraphics[width=75mm, angle=90]{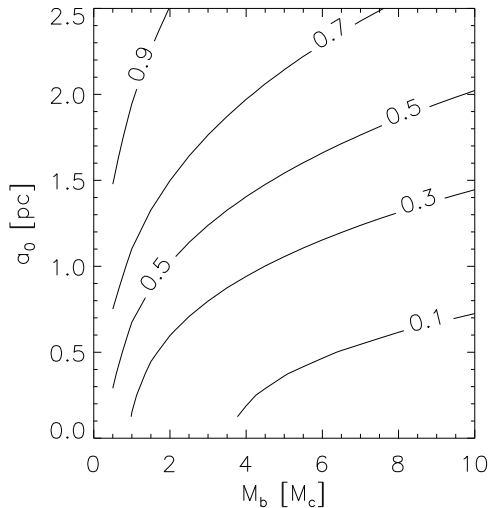}
\caption{
         Contour of $\beta$, the cluster-cloud mass ratio at $r_{40}$ of the clusters.
         Cloud and cluster have the same mass within $r_{40}$ when $\beta=0.5$.
         }
\label{lsfe}
\end{figure}

For later discussions, we introduce the cluster-cloud mass ratio at 40\% Lagrangian radius
of the cluster
\begin{equation}\label{beta_eq}
\beta=\frac{M_{c,r40}}{M_{c,r40}+M_{b,r40}}\,,
\end{equation}
where $M_{b,r40}$ is the mass of the cloud within the initial 40\% Lagrangian radius
of the cluster. $M_{b,r40}$ depends on $a_0$ and $M_b$.
Fig.~\ref{lsfe} shows the contour of $\beta$ in the parameter space $(M_b,a_0)$.
Note that $M_{c,r40}$ is constant (in fact $1000M_\odot$) in our simulations
(as $M_c=2500M_\odot$ in all our simulations).

From its definition, $\beta$ describes star formation efficiency in a certain sense.
In fact, there is an explicit relation between $\beta$ and $\eta$ (the commonly defined
star formation efficiency),
\begin{equation}
\frac{1}{\beta}=1+\frac{r_{40}^3}{M_{c,r40}(r_{40}^2+a^2)^{3/2}}\left(\frac{1}{\eta-1}\right)\,.
\end{equation}

We find that our simulation results on the effect of the parent cloud on the cluster is
better described by $\beta$.
Further discussion will be given later and the 40\% Lagrangian radius of cluster
will be denoted as $r_{40}$ hereafter.

\begin{table}
\caption{Total cloud mass in unit of cluster mass}
\begin{center}
\begin{tabular}{l c c c c c c c c c c}

index  & 01 & 02 & 03 & 04 & 05 & 06 & 07 \\
 {\it M$_b$} [{\it M$_{c}$}] &  0.5 & 1 & 1.5 & 2 & 2.5 & 3 & 3.5\\ \hline
 & 08 & 09 & 10 & 11 & 12 & 13 & 14 \\
 {\it M$_b$} [{\it M$_{c}$}] & 4 & 4.5 & 5 & 5.5 & 6 & 6.5 & 7 \\ \hline
 & 15 & 16 & 17 & 18 & 19 & 20 \\
 {\it M$_b$} [{\it M$_{c}$}] & 7.5 & 8 & 8.5 & 9 & 9.5 & 10 \\

\end{tabular}
\end{center}
\label{para_mass}
\end{table}

\subsubsection{Initial length scale $a_0$}
The length scale of the Plummer potential describes the compactness of the cloud.
The more compact the cloud is the more its effect on the cluster is when it disperses.
We consider $a_0$ from 0.125 pc (compact) to 2.5 pc (loose) as listed in
Table~\ref{para_length}.

\begin{table}
\caption{Initial length scale of the cloud}
\begin{center}
\begin{tabular}{l c c c c c c c c c c}
 index & a & b & c & d & e & f & g \\
 {\it a$_0$} [pc] & 0.125 & 0.25 & 0.375 & 0.5 & 0.625 & 0.75 & 0.875 \\ \hline
 & h & i & j & k & l & m & n \\
 {\it a$_0$} [pc] & 1 &  1.125 & 1.25  & 1.375 & 1.5 & 1.625 & 1.75 \\ \hline
 & o & p & q & r & s & t  \\
 {\it a$_0$} [pc] & 1.875 & 2 & 2.125  & 2.25 & 2.375 & 2.5 \\

\end{tabular}
\end{center}
\label{para_length}
\end{table}

Each set $(\alpha,M_b,a_0)$ represents a model, and we name the model by the indices in
Tables~\ref{para_time}, \ref{para_mass} \& \ref{para_length}.
For example, B10h corresponds to dispersion rate $\alpha=0.2$ (i.e., $t_e$=1.5 Myr),
total cloud mass $M_b=5M_c=12500M_\odot$, and initial length scale of cloud $a_0=1$ pc.


\subsection{Model for the cluster}

\subsubsection{Initial conditions}\label{sec:initial}
\begin{figure}
\includegraphics[width=85mm, angle=0]{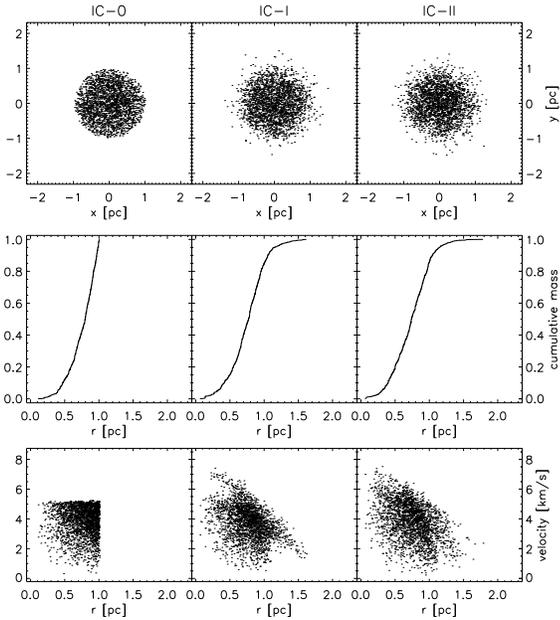}
\caption{
$xy$-projection of the spatial, mass and velocity distributions in radius of stars for
the initial conditions IC-0, IC-I and IC-II, which correspond to preparing the cluster conditions
at 0, 1 and 2 relaxation times, respectively (see text for details).
}
\label{IC}
\end{figure}

The star cluster is prepared
according to a Plummer distribution both in physical
positions and velocities, which are required to achieve virial equilibrium.
Note that the fact the cluster is in virial equilibrium does not mean that it is also
in dynamical equilibrium (see e.g., \cite{Lada1984}, \cite{Goodwin1997}).
In fact, right after we start the simulation, the cluster oscillates
(shrinking and expanding) for a few times before settle down to a smoother
(and more ``natural'') distribution (see Fig.~\ref{IC}).
We, therefore, generate initial conditions according to the following steps:
\begin{itemize}
  \item generate a cluster with a Plummer distribution and a size about 1 pc;
and we called this initial condition IC-0;
  \item put the cluster into a molecular cloud, represented as a Plummer potential,
    with its centre of mass coincides with the centre of the potential;
  \item turn off the cloud dispersion (i.e., $\alpha=0$), and run the code to
two relaxation times; we called the conditions at one and two relaxation times
IC-I and IC-II, respectively.
\end{itemize}
To run the simulation, we turn on the cloud dispersion (i.e., $\alpha\ne 0$),
after zero, one or two relaxation times accordingly.

Fig.~\ref{IC} shows the $xy$-projection of the spatial and velocity distributions
of different initial conditions. For IC-0, stars are restricted within 1 pc, and
the velocity distribution has a very sharp upper limit.
For IC-1 and IC-II, the cluster has smoother spatial and velocity distributions.

\subsubsection{Stellar mass function}\label{sec:massfunc}
\cite{Goodwin1997} mentioned that there is no significant difference between including
a stellar mass function or not.
However, for completeness and for comparison, we consider clusters with and
without mass function.
In both cases, the number of stars is 2500 and the total mass of the cluster is
2500 $M_\odot$.
In cluster with equal mass stars, each star is 1 $M_\odot$.
In cluster with a stellar mass function, we adopt Salpeter mass function with slope -2.35
(\cite{Salpeter1955}) and mass range from 0.32 to 32 M$_\odot$.

\section{Results}

In order to investigate the problem as thoroughly as we can,
for each dispersion rate $\alpha$, each initial condition and each mass function,
we perform four hundred simulation runs on the parameter space $(M_b,a_0)$ as listed in
Tables~\ref{para_mass} \& \ref{para_length}.
We worked out five dispersion rates (see Table~\ref{para_time}), three initial conditions
(see \S\ref{sec:initial}),
and two mass functions equal mass and Salpeter mass funtion, see \S\ref{sec:massfunc}).

To illustrate the main results, we present four typical cases as listed in Table~\ref{modes}:
\begin{itemize}
  \item B14s, $M_b=7.0M_c$, $a_0=2.375$ pc (loose cloud);
  \item B07h, $M_b=3.5M_c$, $a_0=1.0$ pc (intermediate cloud);
  \item B10g, $M_b=5.0M_c$, $a_0=0.875$ pc (intermediate cloud);
  \item B18e, $M_b=9.0M_c$, $a_0=0.625$ pc (massive and compact cloud).
\end{itemize}
All of these cases have $\alpha=0.2$ (i.e., $t_e=1.5$ Myr), with initial condition IC-0 and Salpeter mass function.

\begin{figure}
\includegraphics[width=75mm, angle=90]{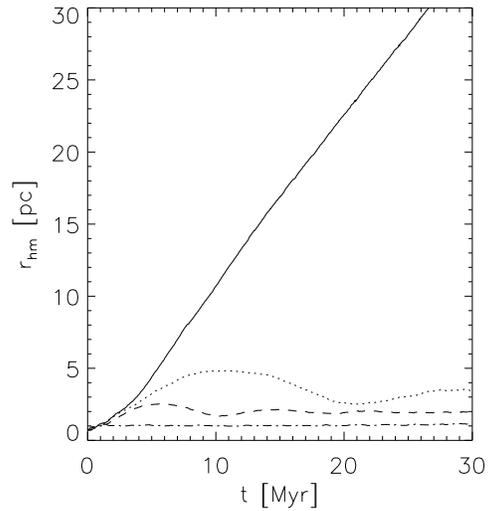}
\caption{
   The evolution of half-mass radius $r_h$. Solid line is the destroyed case B18e.
   Dash-dotted line is the case with a compact core B14s.
   Dashed line and dotted line are loose cases B07h and B10g, respectively.
}
\label{lagr_3cases}
\end{figure}

Fig.~\ref{lagr_3cases} shows how the half mass radii, $r_h$, vary with time
for these four cases. In B14s (dash-dotted line), $r_h$ increases only a little bit.
In B07h (dashed line), $r_h$ increases more than 2 times and becomes stable after 10 Myr.
B10g (dotted line) behaves similar to B07h, but the time to become stable is longer.
Generally, $r_h$ increases in the first 5 to 10 Myr and then shrinks back.
Cluster is considered survived in these three cases (B14s has a compact core, and
B07h and B10g are called loose).
In B18e (solid line), $r_h$ increases almost linearly.
Cases such as B18e, $r_h$ will never decrease again and cluster is considered destroyed.
More quantitative criteria for compact core-, loose- and destroyed-clusters will be given
below.

\begin{table}
\caption{Typical cases}
\begin{center}
\begin{tabular}{c c c c}
Run & {\it t$_e$} & {\it M$_b$} & {\it a$_0$}  \\
 & [Myr] & [{\it M$_{c}$}] & [pc] \\ \hline
B14s & 1.5 & 7 & 2.375 \\
B07h & 1.5 & 3.5 & 1 \\
B18e & 1.5 & 9 & 0.625 \\
B10g & 1.5 & 5 & 0.875 \\ \hline
\end{tabular}
\end{center}
\label{modes}
\end{table}%

\subsection{Effect of cloud dispersion}\label{sec:dispersion}
\begin{figure}
\includegraphics[width=85mm, angle=90]{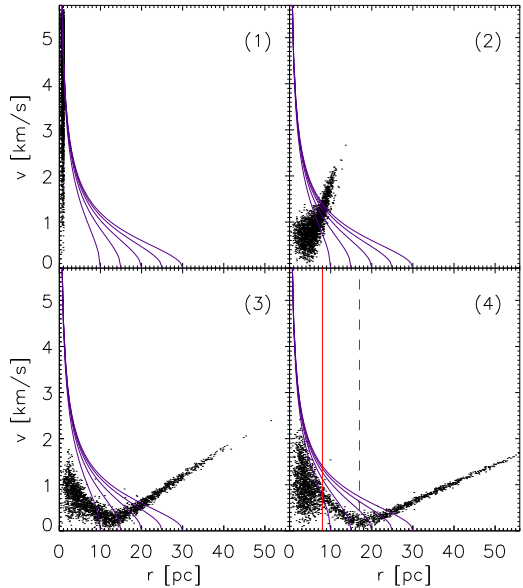}
\caption{
  Different evolutionary stages of case E10i in $\alpha$=0.5 ($t_e$=0.625 Myr).
  Solid lines are tracks of elliptical orbit which semi-major axes are
  10, 15, 20, 25 and 30 pc. There are four stages: (1) bound, (2) expansion,
  (3) inner part formation, and (4) final structure at 30 Myr. Red solid and dash lines in (4) seperate inner part, return stars and stars do not reach their aphelion.
}
\label{e10I_hf}
\end{figure}

The star cluster is in virial equilibrium in the potential well of the molecular cloud
in the beginning.
Every star is moving in a way according to how significant it is bound to the cloud and
other stars.
As the cloud is dispersing, its potential well is getting flatter and flatter,
and its effect is negligible after ~10 Myr.
During this process, the cluster expands. Some stars may escape.
If the expansion is not too serious, the cluster could adjust and return to equilibrium
after some time, otherwise, it is on the way to destruction.
From the velocity distribution, we can describe the evolution of the cluster in four stages as
shown in Fig.~\ref{e10I_hf}. In the figure, the dots are stars, the solid lines are tracks of
elliptical orbits which semi-major axes are 10, 15, 20, 25 and 30 pc.
\begin{itemize}
\item[(1)] bound: stars orbiting the centre of mass of the cluster at
high speed because of the additional gravity provided by the cloud;
\item[(2)] expansion: during (and after) cloud dispersion, stars seem to escape;
\item[(3)] inner part formation: stars with shorter periods may pass their aphelion and return
(following the solid lines which describe the Keplerian orbits);
\item[(4)] final structure at 30 Mry: stars with longer periods may return, and return stars will forget their previous orbits
due to the complex interaction in the inner part.
\end{itemize}

\begin{figure}
\includegraphics[width=85mm, angle=0] {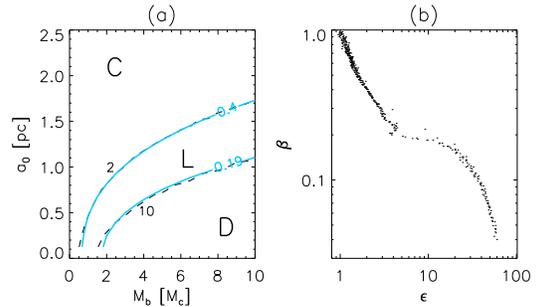}
\caption{
  Groups in the cases with $t_{e}$= 1.5 Myr and Salpeter mass function.
  (a) Three generic groups: destroyed (D), loose (L), and compact core (C) are shown in
  the parameter space $(M_b,a_0)$. The dashed lines are the expansion ratio $\epsilon=2$ and 10.
  The solid lines are cluster-cloud mass ratio $\beta=0.40$ and 0.19, respectively.
  The two sets resemble each other very well (they almost overlap with each other).
  (b) Relation between $\epsilon$ and $\beta$. Each points on the figure is a simulation run
  (there are 400 points).
  It appears linear for $\epsilon<5$, there is a lack of data in the interval $5<\epsilon<10$
  and for $\epsilon>10$ some other relation takes over.
}
\label{threepart}
\end{figure}

\begin{figure}
\includegraphics[width=75mm, angle=0]{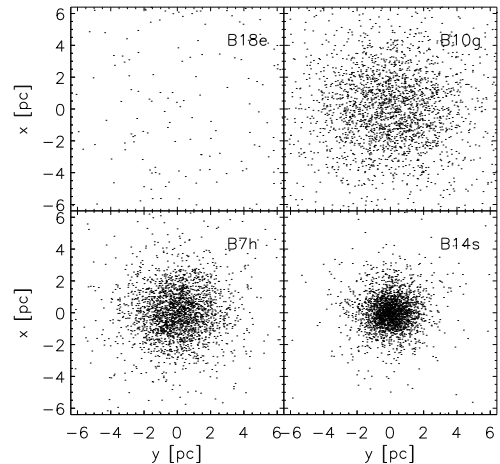}
\caption{
         $xy$-projection of the spatial distribution of the cluster at 30 Myr.
         There are three generic types: destroyed (B18e), loose (B07h \& B10g),
         compact core (B14s).
         }
\label{xyfor3cases}
\end{figure}

However, not every star could return if the dispersion of cloud is too serious.
Our results show that the final structure can be grouped into three groups:
(a) destroyed, (b) loose, and (c) compact core.
Fig.~\ref{threepart} shows the three groups in the parameter space $(M_b,a_0)$.
By examining the spatial distribution, we find that the boundaries can be well described
by a parameter $\epsilon$ which we called the expansion ratio.
It is defined as the ratio between the final and initial $r_{40}$ of the cluster,
\begin{equation}\label{epsilon_eq}
\epsilon=\frac{\it r_{40,f}}{\it r_{40,i}}\,.
\end{equation}
The boundaries are $\epsilon\approx 2$ and 10.
These numbers are chosen naively by examing the spatial distribution by eyes.
\begin{itemize}
\item[(a)] destroyed ($\epsilon>10$): when clusters could not adjust themselves back
to equilibrium in time, they will expand forever; e.g., case B18e;
\item[(b)] loose ($10>\epsilon>2$): most clusters could adjust back to equilibrium but may become looser and loose part of stars;
e.g., cases B07h and B10g;
\item[(c)] compact core ($2>\epsilon$): in some cases, $r_{40}$ expands slightly
(less than twice), they become stable and develop a denser, e.g. case B14s.
\end{itemize}
Fig.~\ref{xyfor3cases} shows the $xy$-projection of the final spatial distribution of
the cluster of the cases B18E, B07h, B10g and B14s.
Note that they have the same cloud dispersion rate.

\subsection{Cluster-cloud mass ratio and expansion ratio at $r_{40}$}\label{sec:ratio}
As the cluster is bound by the parent cloud and its member stars initially.
One would expect the more compact the cloud is the more the cluster will expand
when the cloud is dispersed.
After some experimentation, we find that the best parameters to delineate this
anti-correlation is the initial cloud-cluster mass ratio $\beta$ and the
expansion ratio $\epsilon$, both define at the 40\% Lagrangian radius of the
cluster $r_{40}$.
Fig.~\ref{threepart}b shows how tight the correlation is.
We should point out that if we define the cloud-cluster mass ratio and the expansion ratio
with respect to other Lagrangian radii (say, 30\% to 50\%),
we get qualitatively similar results but the
correlation is not as good as the 40\% Lagrangian radius.
We still do not understand what is so special about $r_{40}$ in our system.

\begin{figure}
\includegraphics[width=45mm, angle=90]  {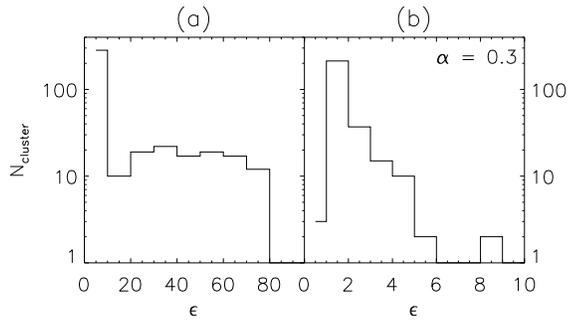}
\caption{Histogram of $\epsilon$ for $t_e$=1.1 Myr.
        (a) Histogram from $\epsilon$=0 to 100. About 300 cases (out of a total 400)
        are under 10.
        (b) Histogram from $\epsilon$=0 to 10. Most are under 2 and the number
         drops off as $\epsilon$ increases. }
\label{eps_imf0.3_histo}
\end{figure}

It is worthwhile to mention the statistics of $\epsilon$ in our simulation sample
(with same dispersion rate and mass function).
A typical result is shown in Fig.~\ref{eps_imf0.3_histo}, which shows the histogram
of the number of runs resulting in $\epsilon$.
Clearly, most of the runs lie in $\epsilon<10$ and
for $\epsilon>10$ the data seems to be evenly distributed (Fig.~\ref{eps_imf0.3_histo}a).
When we zoom into $0<\epsilon<10$, we find that there is a drop off in $2<\epsilon<5$,
and within $5<\epsilon<10$ there is almost no data (Fig.~\ref{eps_imf0.3_histo}b).
It seems that there is a transition region between 5 to 10.

\subsection{Dispersion rates}\label{sec:rates}
\begin{figure}
\includegraphics[width=75mm, angle=0] {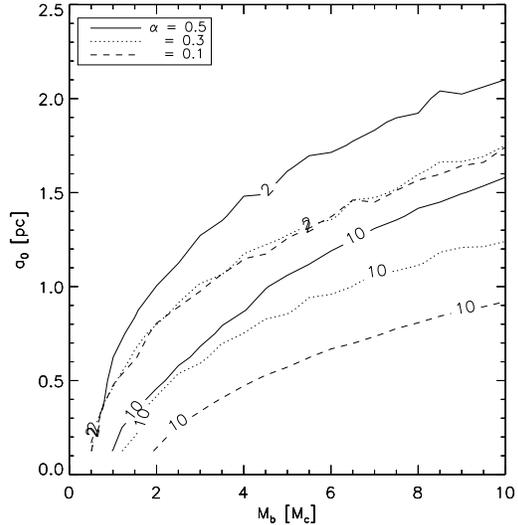}
\caption{
  The three generic groups in different dispersion rate (cf, Fig.~\ref{threepart}a).
  Solid line is for $\alpha$=0.5 ( $t_{e}$=0.625 Myr),
  dotted line is for $\alpha$=0.3 ( $t_{e}$=1.1 Myr) and dashed
  line is for $\alpha$=0.1 ($t_{e}$=3.3 Myr). For $\alpha$=0.1 and 0.3,
  the $\epsilon\approx 2$ lines almost overlap, and although there is a separation in
  the $\epsilon=10$ lines, they just look like a shifting. For $\alpha$=0.5,
  the most rapid dispersion, although it looks a little different than the other two,
  we can still divide it to the three generic groups.
}
\label{threepart_c_imf.tr0}
\end{figure}

We work out five dispersion rates of the cloud with e-fold time $t_{e}$ is from
0.625 to 3.3 Myr (see Table~\ref{para_time}).
Fig.~\ref{threepart_c_imf.tr0} shows the $r_{40}$ expanding ratio $\epsilon$ of
three dispersion rates (with e-fold time $t_e=$ 3.3, 1.1, 0.625 Myr)
with the same initial condition (IC-0).
The $\epsilon\approx 2$ lines of the cases $t_{e}=$ 3.3 and 1.1 Myr almost overlap.
As expected the region for destroyed cluster ($\epsilon>10$) is larger for
$t_{e}$=0.625 Myr (the most rapid dispersion rate in our simulations).
In any case, many (more than two-third) of the clusters survive and remain intact
30 Myr after the cloud started to disperse.

\begin{figure}
\includegraphics[width=75mm, angle=90]{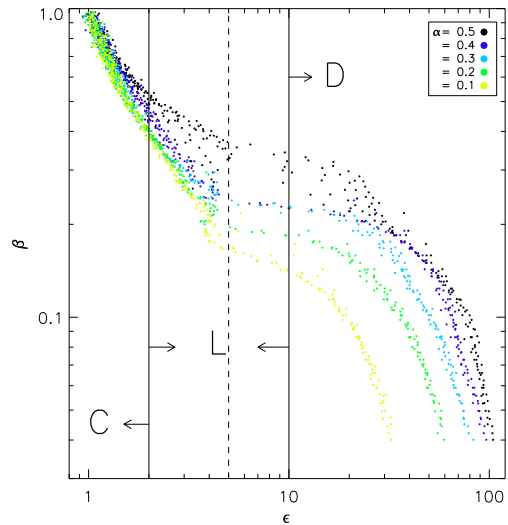}
\caption{
         Relation between $\epsilon$ and $\beta$ for different dispersion rates.
         The relations are fairly similar for the cases $\alpha=$ 0.1, 0.2, 0.3 and 0.4.
         It is kind of different for $\alpha=$ 0.5, but we can still use the
         solid lines $\epsilon=2$ and 10 to separate the destroyed, loose and compact
         core groups.
}
\label{imf_eps_all}
\end{figure}

Fig.~\ref{imf_eps_all} shows the relation between $\epsilon$ and $\beta$ for all
five dispersion rates. Except $t_e$=0.625 ($\alpha$=0.5), all of them
look alike when $\epsilon<5$ (dashed line is $\epsilon=5$).
Even for $t_e$=0.625, the trend ($\beta$ is larger when $\epsilon$ is smaller)
is the same as the others.

From Fig.~\ref{imf_eps_all} we can read the corresponding value of $\beta$ for
$\epsilon=$ 2 and 10 for the five dispersion rates. The result is listed in
Table~\ref{lsfe_result}.

\begin{table}
\caption{The value of $\beta$ corresponding to $\epsilon=2$ and 10.}
\begin{center}
\begin{tabular}{l c c c c l}
t$_e$  [Myr] & 3.3 & 1.5 & 1.1 & 0.75 & 0.625 \\ \hline
$\beta_2$ ($\beta$ at $\epsilon=2$) & 0.40 & 0.40 & 0.40 & 0.45 & 0.50 \\
$\beta_{10}$ ($\beta$ at $\epsilon=2$) & 0.13 & 0.19 & 0.22 & 0.22 & 0.30 \\
\end{tabular}
\end{center}
\label{lsfe_result}
\end{table}%

\begin{figure}
\includegraphics[width=75mm, angle=90] {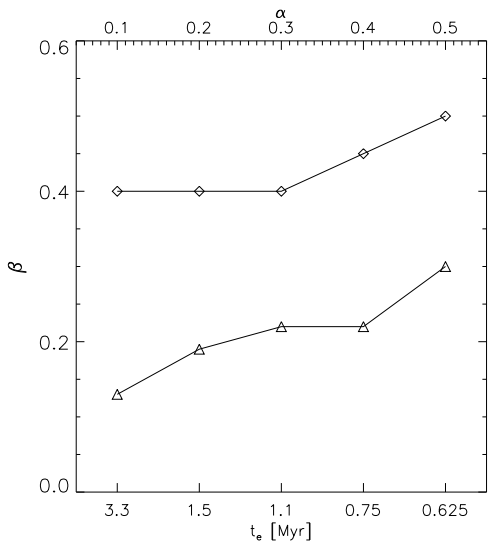}
\caption{
The relation between $\alpha$ and $\beta$ (or $t_{e}$ and $\beta$) for $\epsilon=2$
(diamond and the upper line) and 10 (triangle and the lower line).
}
\label{beta_im}
\end{figure}

\subsection{Initial conditions in virial equilibrium}

Similar works usually start the cluster from virial equilibrium state.
However cluster in virial equilibrium might not necessary in dynamical equilibrium
(e.g, \cite{Goodwin1997} and our tests,
a brief description of our tests is given in \S\ref{sec:initial}).
We would like to know whether this would make a difference.
Thus we consider three different initial conditions according to the prescription
mentioned in \S\ref{sec:initial} (IC-0, IC-1, IC-2 for conditions taken at 0, 1, 2
relaxation times).

\begin{figure}
\includegraphics[width=75mm, angle=0] {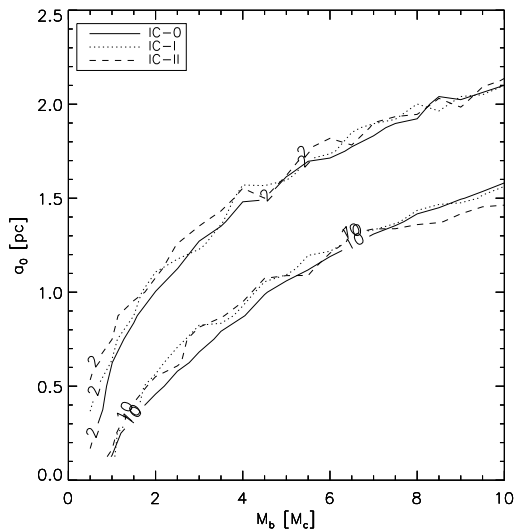}
\caption{
         The three generic groups for $t_e=$ 0.625 Myr in different initial conditions.
         Solid line, dotted line and dashed line are for IC-0, IC-I and IC-II, respectively.
         There is no significant difference between these three models. Nonetheless
         IC-0 has the smoothest boundaries.
}
\label{threepart_c_imf.alpha0.5}
\end{figure}

Fig.~\ref{threepart_c_imf.alpha0.5} shows the results of $t_e=$ 0.625 Myr for different
initial conditions.
The results agree with each other reasonably well. Notheless IC-0 gives the smoothest divisions.
Similar results are seen in other models.

\subsection{Stellar mass function and mass segregation}
Previous works, such as \cite{Goodwin2006,Bastian2006}, claimed that simulations with
equal mass and with mass function should give similar outcome.
For completeness, we re-examine this issue.
All the results presented so far are cases with Salpeter mass function (see \S\ref{sec:massfunc}).
We repeat all the simulations for clusters with equal mass (i.e., each star is 1 $M_\odot$).
Note that the number of stars and the total mass of the cluster are the same as
those clusters with mass function (see \S\ref{sec:massfunc}).

\begin{figure}
\includegraphics[width=75mm, angle=0]{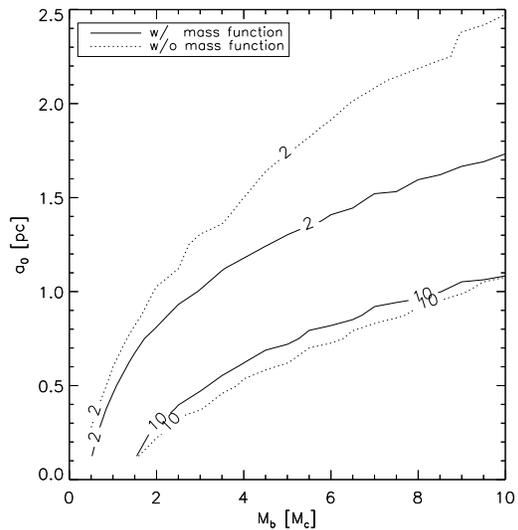}
\caption{
The three generic groups in different mass functions.
The dispersion e-fold time of the cloud is $t_e=$ 1.5 Myr.
Solid line is clusters with Salpeter mass function
and dotted line is clusters with equal mass.
The lines of $\epsilon=10$ agree well with each other,
but the lines of $\epsilon=2$ are quite different,
which is supposed to be a result of mass segregation.
}
\label{threepart_c_both.alpha0.2}
\end{figure}

The three generic groups: destroyed, loose and compact core are still applicable
in the case of equal mass clusters.
Fig.~\ref{threepart_c_both.alpha0.2} compares the results with and without
mass function in the same dispersion rate and the same initial condition (IC-0).
In the figure, the solid lines are clusters with Salpeter mass function and the
dotted lines are clusters with equal mass.
While the boundaries seperating the loose group and destroyed group for the two mass functions
agree well with each other, the lines seperating the loose group and compact core group do not.
This is true for every dispersion rate.
We conclude that only the boundary between loose group and destroyed group does
not depend on mass function.
The boundary between loose group and compact core group are different and may be
attributed to mass segregation.

\begin{figure}
\includegraphics[width=75mm, angle=90]{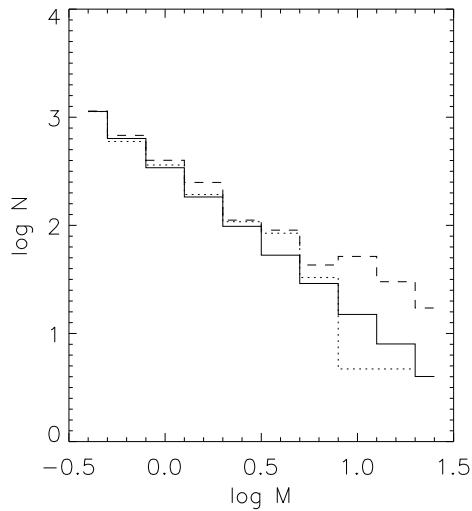}
\caption{
  Mass segregation of case C14s at 30 Myr (the end of our simulation).
  Solid line, dashed line, dotted line are the mass functions for
  the whole cluster (Salpeter), for stars within 1 pc (shallower than Salpeter),
  for stars between 1 pc to 1.5 pc (steeper than Salpeter), respectively.
  Note that all the first bins are scaled to the same number.
}
\label{mdist}
\end{figure}

The concentration ratio for surviving clusters should be affected by mass function.
Even for lower cluster-cloud mass ratio $\beta$ could form a dense core in model
with mass function.
Examining mass distribution at later time, one should find that there is (and should be)
mass segregation for cases with mass function.
Fig.~\ref{mdist} shows the mass function of case C14s at 30 Myr (the end of our simulation).
Since we do not have stellar evolution, there is no mass loss during the simulation.
The mass function for the whole cluster remains Salpeter (solid line).
In the inner region (1 pc, dashed line) the mass function is shallower, while in the
outer region (between 1 to 1.5 pc, dotted line) the mass function is steeper or at least
the region has less massive stars.

\begin{figure}
\includegraphics[width=75mm, angle=90] {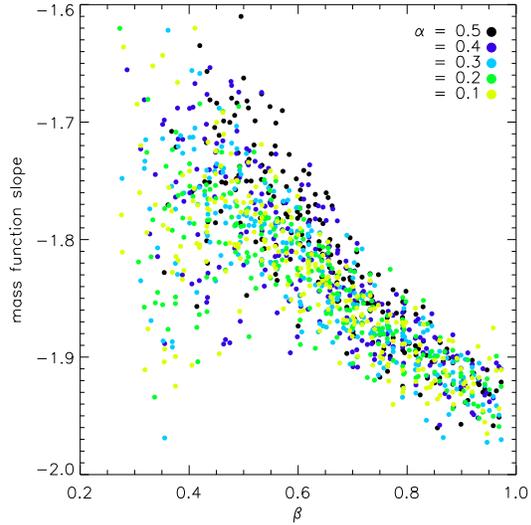}
\caption{
         Slope of mass function (within initial $r_h$) against cluster-cloud mass ratio for different dispersion
         rates. The mass function of initial condition is Salpeter with slope -2.35.
         The general trend is more serious mass segregation when $\beta$ is smaller.
         The scattering in small $\beta$ might be due to the number of stars is small
         at 30 Myr (the end of simulation).
}
\label{mfs_tr0}
\end{figure}

\begin{figure}
\includegraphics[width=75mm, angle=90] {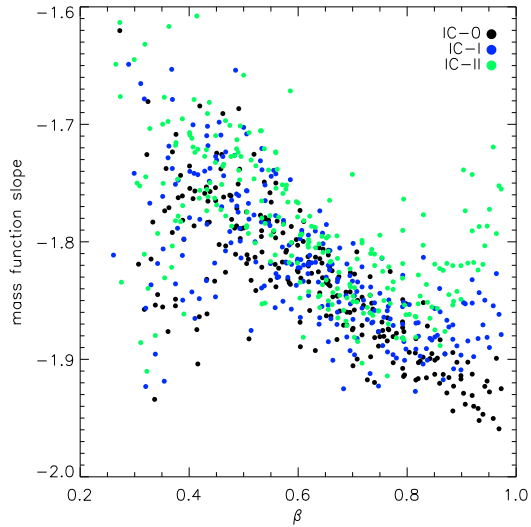}
\caption{
         Slope of mass function (within initial $r_h$) against cluster-cloud mass ratio
         for the three initial conditions (\S\ref{sec:initial}).
         The cloud dispersion e-fold time is $t_e=1.5$.
         The general trend is more serious mass segregation when $\beta$ is smaller.
         However, when $\beta\geq 0.8$ there is another increase in mass segregation for
         IC-I and IC-II (see text for a possible explanation).
}
\label{mfs_a0.2}
\end{figure}

To learn how dispersion rates and initial conditions affect mass segregation,
we plot the slope of the mass function against the cluster-cloud mass ratio $\beta$
(Figs.~\ref{mfs_tr0} \& \ref{mfs_a0.2}).
The general trend is the mass function is shallower when $\beta$ is smaller,
i.e., mass segregation is more serious at smaller $\beta$.
This can be understood as less massive stars are more likely to escape.
(Recall that the cluster-cloud mass ratio $\beta$ and the $r_{40}$ expansion ratio
$\epsilon$ are anti-correlated, see Fig.~\ref{imf_eps_all} or \ref{threepart}b.)

In Fig.~\ref{mfs_tr0}, there is little difference between different
dispersion rates when $\beta>0.5$.
For smaller $\beta$ (i.e., when cloud dominates over cluster in the beginning),
the data scatter a lot.
The reason might be the number of stars remains in the cluster is small in these cases,
and the statistics is not very good.

In Fig.~\ref{mfs_a0.2}, besides more serious mass segregation at small $\beta$,
there is another increase in mass segregation when $\beta \geq 0.8$ for IC-I and IC-II.
The reason can be traced back to the preparation of the initial conditions.
For IC-I and IC-II, the cluster has already had some mass segregation in the beginning.
It segregates more when $\beta$ is large. After cloud dispersion,
it becomes even more segregated.

\subsection{Infant mortality}
It is conceivable that the dispersion of a molecular cloud may destroy its embedded
infant star clusters. The question is how effective is the process.
From our simulation results, we deem that the process is not very effective.
Infant clusters can be destroyed only when the $r_{40}$ cluster-cloud mass ratio
$\beta$ is below some value. Specifically, $\beta<\beta_{10}$ ($\beta$ at $\epsilon=10$)
in Table~\ref{lsfe_result}
(also see Fig.~\ref{beta_im}).
From the cases we considered, many infant clusters survive (some of them are loosened
while some develop a compact core).
Table~\ref{IM} presents the frequency of the three groups (destroyed, loose, compact core)
for different cloud dispersion rates and mass functions.
Roughly 20\% to 40\% (e-fold time from 3.3 Myr to 0.625 Myr) will be destroyed.
More than 30\% for equal mass clusters and more than 40\% for Salpeter mass function
develop a compact core.

We should point out that the boundaries ($\epsilon=$ 2 and 10)
between the three groups compact core, loose and destroyed are determined
partly by eyes (from the spatial, velocity and mass distribution such as in
Fig.~\ref{xyfor3cases}) and
partly by numbers (such as the histograms in Fig.~\ref{eps_imf0.3_histo}).
The boundaries between loose and destroyed ($\epsilon=10$) we pick may not be precise.
However, we claim that the exact location of the boundary (if there is an extra one)
does not affect the estimated destroyed frequency listed in Table~\ref{IM}.
As we discussed in \S\ref{sec:ratio} there is a transition zone between
$\epsilon=$ 5 to 10 where only very few cases exist (see Fig.~\ref{eps_imf0.3_histo}).
As the boundary is somewhere around $\epsilon\approx 10$ (within the transition zone),
the exact location will not matter too much on the estimated frequency for the two
classes (destroyed and loose).

We should point out that the range of the transition region is actually slightly different
for different dispersion rates (see Fig.~\ref{imf_eps_all}).
In most cases, the transition region runs from $\epsilon=$ 5 to 15 (for slow dispersion,
such as $t_e=$ 3.3 Myr) and 5 to 30 (for rapid dispersion, such as $t_e=$ 0.625 Myr).
The exact location does not matter too much on the results as there are only very few
cases inside the transition region.

\begin{table}
\caption{Infant mortality in \%}
\begin{center}
\begin{tabular}{l l l l l l}
Salpeter\\ \hline
$t_{e}$ ]Myr]  & 3.3 & 1.5 & 1.1 & 0.75 & 0.625 \\ \hline
Compact core & 55 & 54 & 54 & 49.25 & 42.75 \\
Loose & 26 & 22 & 16.75 & 21.75 & 21\\
Destroyed & 19 & 24 & 29.25 & 29 & 36.25\\

\vspace{2mm}\\
equal mass\\ \hline
$t_{e}$ [Myr] & 3.3 & 1.5 & 1.1 & 0.75 & 0.625 \\ \hline
Compact core & 37.5 & 36.5 & 35.5 & 32.5 & 30 \\
Loose & 45 & 42.25 & 38.5 & 38 & 37.75\\
Destroyed & 17.5 & 21.25 & 26 & 29.5 & 32.25\\
\end{tabular}
\end{center}
\label{IM}
\end{table}%

\section{Summary and remarks}
Can a bound star cluster in a molecular cloud survive when the parent cloud is dispersed?
We studied this problem by means of N-body simulations.
The dispersion of the cloud makes the cluster to expand or even dissociate as the binding
energy decreases.
However, our simulations show that this process is ineffective in destroying the
bound cluster.

The dispersion of the cloud is modelled by a Plummer potential with expanding length scale.
We numerically simulated the behaviour of a bound cluster in this potential up to 30 Myr.
We performed numerous simulations and concluded that there are three groups of final
morphology: compact core, loose and destroyed.
It turns out that the 40\% Lagrangian radius expansion ratio of the cluster $\epsilon$
(the ratio of $r_{40}$ at 30 Myr to initial, see Eq.~(\ref{epsilon_eq}))
is a good indicator for the final fate of the cluster.
More importantly, we found that the initial $r_{40}$ cluster-cloud mass ratio $\beta$
(see Eq.~(\ref{beta_eq})) is tightly correlated with $\epsilon$.
Hence we can ``predict'' the fate of the embedded bound cluster by examining its $\beta$.

The survival probabilities are high when $\beta$ is high.
After 30 Myr, more than 70\% of the clusters still remain a shape of clusters, with a
number density higher than 15/pc$^3$ within 1 pc radius.
Different cloud dispersion rates provide similar results and even
the largest rate in our simulations (e-fold time $t_e=0.625$ Myr) does not disrupt all the
clusters.
Systems with and without mass function have different final densities but agree
with each other well.

We conclude that the infant mortality rate should be low if the embedded clusters are
bound from the beginning, unless $\beta$ is less than about 0.13
(for slow dispersion, such as $t_e=3.3$ Myr) to 0.30 (for rapid dispersion, such as $t_e=0.625$ Myr)
(see Table~\ref{lsfe_result}).

Near infrared observations indicate that the survival probability for embedded clusters
is about 4\% to 7\% (\cite{Lada2003}).
How should we reconcile our result with the observations?
We deem that the dispersion of parent cloud can not account for the destruction of
embedded bound clusters.
Therefore, other more effective mechanisms must be responsible or the stars formed in
molecular cloud are not bound (see e.g., \cite{Bonnell2006}).

\section*{Acknowledgments}
We appreciate Alessia Gualandris for her kindly and very helpful discussions.
This work was supported in part by the National
Science Council, Taiwan under the grants
NSC-95-2112-M-008-006 and NSC-96-2112-M-008-014-MY3.


\end{document}